\documentclass[twocolumn, trackchanges]{aastex631}
\usepackage{amsmath,amssymb,gensymb,times,graphicx,morefloats,color,wrapfig}

\newcommand{\gaia}{Gaia}

\newcommand{\kepler}{{\it Kepler}}

\newcommand{\teff}{\ensuremath{T_{\text{eff}}}}

\newcommand{\tess}{\textit{TESS}}
\newcommand{\ktwo}{{\textit K2}}

\newcommand{\prot}{$P_{\rm{rot}}$}
\newcommand{\Prot}{P_{\rm{rot}}}
\newcommand{\protCPM}{$P_{\mathrm{rot, \, CPM}}$}
\newcommand{\protRH}{$P_{\mathrm{rot, \, RH20}}$}

\graphicspath{{Figures/}{../Figures/}{Figs_old/}{.}}

\shorttitle{Completeness and Reliability of \tess\ stellar rotation}
\shortauthors{Boyle et al.}

\begin{document}

\title{Quantifying the Limits of \tess\ Stellar Rotation Measurements with the \ktwo-\tess\ Overlap}

\author[0000-0001-6037-2971]{Andrew W. Boyle}
\altaffiliation{NSF Graduate Research Fellow}
\affiliation{Department of Physics and Astronomy, The University of North Carolina at Chapel Hill, Chapel Hill, NC 27599, USA}

\author[0000-0003-3654-1602]{Andrew W. Mann}
\affiliation{Department of Physics and Astronomy, The University of North Carolina at Chapel Hill, Chapel Hill, NC 27599, USA}

\author[0000-0002-9446-9250]{Jonathan Bush}
\affiliation{Department of Physics and Astronomy, The University of North Carolina at Chapel Hill, Chapel Hill, NC 27599, USA}

\correspondingauthor{Andrew Boyle}
\email{awboyle@unc.edu}

\begin{abstract}

The Transiting Exoplanet Survey Satellite (\tess) has provided stellar rotation periods across much of the sky through high-precision light curves, but the reliability and completeness of these measurements require careful evaluation. We assess the accuracy of \tess-derived rotation periods by leveraging a cross-matched sample of $\sim$23,000 stars observed by both \tess\ and the \ktwo\ mission, treating \ktwo\ periods as a benchmark. Using causal pixel models to extract light curves and the Lomb-Scargle (LS) periodogram to identify rotation signals, we quantify the empirical uncertainties, reliability, and completeness of \tess\ rotation period measurements. We find that uncertainties on \tess-derived rotation periods are typically below 3\% for stars with periods $<$ 10 days. Rotation periods are generally reliable out to 10 days, with $\gtrsim$80\% of measurements matching the \ktwo\ benchmark. Completeness and reliability drop dramatically for periods beyond $\simeq12$\,days due to the 27-day sector limitation. Stricter cuts on \tess\ magnitude and LS power improve reliability; the highest LS power tested ($>0.2$) ensures $>90\%$ reliability below 10\,days but removes over half of potential detections. Stitching consecutive-sector light curves reduces period uncertainties but does not improve overall reliability or completeness due to persistent systematics. Our findings and code provide a framework for interpreting \tess-derived rotation periods and inform the selection of quality cuts to optimize studies of stellar rotation, young associations, and gyrochronology.


\end{abstract}

\section{Introduction} \label{sec:intro}

NASA’s \kepler\ mission \citep{Borucki2010} transformed our understanding of stellar rotation, with light curves that allowed measurement of rotation periods for tens of thousands of stars, extending up to $\simeq$100 days \citep[e.g.,][]{2014ApJS..211...24M, Angus2015, Santos2021}. Building on \kepler's success, the repurposed \ktwo\ mission \citep{VanCleve2016} targeted young stellar populations along the ecliptic plane, enabling the development of gyrochronological sequences for stellar associations with ages from around 10 Myr to over 2 Gyr \citep[e.g.,][]{Rebull2018, Douglas:2019, Curtis2020}.

 NASA's Transiting Exoplanet Survey Satellite \citep[\tess;][]{2014SPIE.9143E..20R} now offers an opportunity to expand stellar rotation studies with its near-complete sky coverage and $<$1\% photometric precision down to $T\simeq15$ \citep{2015JATIS...1a4003R}. \tess\ provides variability metrics for a diverse set of stars, including those of scientific interest such as stars with spectroscopic monitoring \citep[e.g.,][]{Hojjatpanah2020}, wide binaries \citep[e.g.,][]{2016MNRAS.455.4212D}, eclipsing systems \citep[e.g.,][]{Prsa2022}, and M dwarfs \citep[e.g.,][]{Pass2022}.
 
\tess\ has already significantly advanced our knowledge of stellar rotation. Rotation periods derived from \tess\ light curves have been instrumental in confirming diffuse coeval populations \citep[e.g.,][]{THYMEV, Wood2022}, estimating ages of exoplanet-hosting stars \citep[e.g.,][]{Zhou2021}, and investigating properties of nearby brown dwarfs \citep{Apai2021}. As \tess\ continues observing, covering new sky regions and extending time baselines, the scope of rotation-based research in stellar and planetary sciences will continue to grow.

A major challenge with \tess\ data is the 27.4-day observing window, which limits detection of rotation periods beyond $\simeq$13 days within a single sector. While multiple sectors (particularly in \tess’s continuous viewing zones) theoretically provide longer baselines for period measurement, efforts to use these longer baselines have generally struggled to recover reliable long-period rotations \citep[e.g.,][]{VanderPlas2018, CantoMartins2020}. Some progress has been made using machine learning techniques \citep[e.g.,][]{Lu2020, Claytor_2024} or combining \tess\ data with ground-based observations \citep{Howard2021}.

Additional challenges in detecting rotation signals from \tess\ data include scattered light contamination \citep{Hedges2020}, \tess's relatively small aperture for faint stars, \tess’s large pixel scale (21\arcsec) and contamination from nearby stars, and data gaps during each sector’s data downlink period. These factors can hinder period detection, and in some cases, as with scattered light interference, may even introduce erroneous periodic signals \citep[e.g., the $\simeq$13.7-day lunar-synchronous signal;][]{hattoriUnpopularPackageDatadriven2021}.

Given the ubiquity of \tess\ light curves and the potential limitations in deriving accurate rotation periods, it is essential to answer two key questions: how reliable are \tess-derived rotation periods, and, when periods are undetectable, what can be inferred about the star’s rotation? For the former, we assess \textit{reliability}—the accuracy of detected rotation periods. For the latter, we evaluate \textit{completeness}—the thresholds below which rotation signals cannot be reliably detected. 

Both reliability and completeness vary with observable factors such as rotation period and stellar brightness. With a robust map of both as a function of common observables, we can more accurately compare theoretical rotation period distributions, such as those from a single-aged \prot-color sequence, with observed distributions. Similar methods are used in exoplanet research, where knowledge of both false-positive rates and pipeline completeness is crucial for deriving occurrence rates \citep{Thompson2018}. While injection-recovery tests are common in exoplanet studies, simulating rotation signals is more challenging as there is no analytical model that describes stellar rotation comparable to that of transits. The best alternative is to compare rotation periods measured from \tess\ to that of a more reliable/complete set, like those from \kepler\ or \ktwo. 

This study aims to chart the completeness and reliability of rotation periods derived from \tess\ light curves using widely accessible, commonly adopted methods. In the long term, we aim to translate \tess-based rotation measurements and non-detections into meaningful indicators of stellar age or membership in young associations. We therefore seek an extraction method that preserves variability signals even for faint stars in nearby young clusters, as well as a straightforward, interpretable algorithm for periodic signal measurement that can be efficiently applied across large stellar samples.

The rest of the paper is organized in the following fashion: Section \ref{sec:sample} describes the sample of benchmark rotation periods used as to quantify the precision, reliability, and completeness of the \tess\ rotation pipeline. Section \ref{sec:prot} describes the method for extracting light-curves and measuring \prot. Section \ref{subsec:empirical_uncertainties} quantifies the empirical uncertainties of \tess\ \prot\ measurements, while Sections \ref{sec:rel} and \ref{sec:comp} explore the reliability and completeness of single-sector \prot\ measurements, respectively. In Section \ref{sub:multi_results}, we discuss the reliability and completeness of \prot\ measurements from stitched consecutive-sector light curves. We show an application of our results in Section \ref{sec:application} by applying it to a set of nearby stellar clusters. We discuss some of the implications of our work in Section~\ref{sec:discussion} and list the key takeaways for our study alongside future improvements in Section~\ref{sec:conclusion}.

\section{Sample Selection}\label{sec:sample}

To detect the limits of \tess\ \prot\ measurements, we required a large set of stars in the \tess\ field of view with reliable rotation periods (the benchmark set). To this end, it was important to include both stars with longer rotation periods and those that are fainter than the limits we expect for targets observed by \tess. That is, we want to detect the limits (in \prot\ and $T$) where \tess\ fails to recover rotation periods, which requires targets landing beyond those limits. 

\tess\ Cycles 4 and 5 cover the ecliptic plane, offering a large sample of stars with both \tess\ and \ktwo\ data (in addition to earlier \tess\ cycles with some \ktwo\ overlap). \ktwo-\tess\ overlap studies have performed extensive searches of the \ktwo\ data for rotation periods \citep[e.g.,][]{douglas_k2_2016,  douglas_poking_2017, rebull_rotation_2018}. Importantly, \ktwo\ campaigns were significantly longer than a \tess\ sector ($>$70\,days vs $\simeq27$\,days), and \ktwo\ could reach fainter sources than \tess\ (\ktwo's aperture is 95cm vs. 10cm for \tess). Comparisons with ground-based data and across multiple \ktwo\ campaigns have shown \ktwo\ periods to be reliable \citep{rampalliThreeK2Campaigns2021}. Thus, rotation periods derived from \ktwo\ data can serve as a reasonable benchmark to test the reliability of \tess\ \prot\ measurements. 

We used the \ktwo\ rotation periods from \citep[][hereafter RH20]{reinholdStellarRotationPeriods2020}. RH20 performed a comprehensive search for \prot\ between 1 and 44 days across all available \ktwo\ light curves, reporting an empirically estimated error \textit{$e_{Prot}$} and a signal strength metric, \textit{HPeak}, for each \prot. To select stars from this catalog, we adopted RH20’s quality criterion of $\rm{HPeak} > 0.3$ and limited our analysis to $\Prot < 40$ days. Extending our sample in period yielded only a negligible increase in the number of stars. RH20 suggests greater confidence for stars with $\rm{HPeak} > 0.5$, but our results were consistent across the extended range $0.3<\rm{HPeak}<0.5$. Additionally, variations in spot coverage over time and the wavelength differences between \ktwo\ and \tess\ make \rm{HPeak} only a rough proxy for expected signals in \tess\ data. 

To identify stars from the RH20 catalog within the \tess\ field of view, we matched EPIC catalog coordinates for each star to the \texttt{tess-point} software \citep{2020ascl.soft03001B}, determining which stars from RH20 have been observed by \tess. This process provided a sample of approximately 23,000 stars observed by both \tess\ and \ktwo\ as of January 2024.

\begin{figure*}[t!]
    \centering
    \includegraphics[width = 0.97\textwidth]{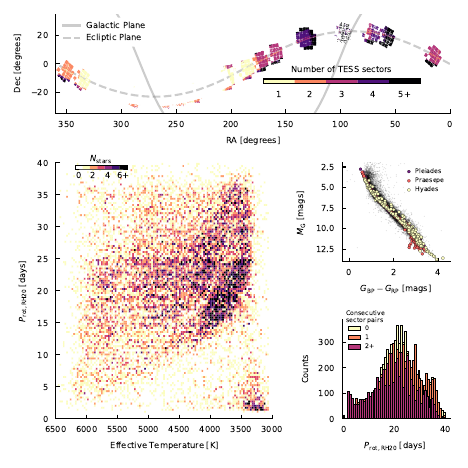}
    \caption{\textbf{The RH20-\tess\ overlap.} Characteristics of the 23,000-star RH20-\tess\ overlap sample used in this study. On top, we show the sky position of targets colored by the number of \tess\ sectors available as of January 2024. The bottom left panel shows the rotation period  distribution (as measured by \ktwo) for the subset of stars that comprise our final sample used for analysis (cf. Section \ref{subsec:quality_cuts}). Effective temperatures are calclulated with the empirical color-effective temperature relation from \cite{curtis_when_2020} with Gaia DR2 $G_{\rm BP} - G_{\rm RP}$ colors dereddened using STILISM dust maps \citep{2019A&A...625A.135L}. The middle-right panel shows the target's position on a \gaia\ color-magnitude diagram, with members of the three major clusters over plotted. The bottom right shows a (stacked) histogram of the RH20 rotation periods colored by number of consecutive \tess\ sectors. The sample covers a wide range in stellar type, rotation period, brightness, and data coverage -- ideal for testing completeness and reliability with \tess. }
    \label{fig:RH20_sample}
\end{figure*}

As we show in Figure~\ref{fig:RH20_sample}, the resulting RH20-\tess\ overlap sample is diverse, spanning a wide range of color (\teff), rotation period, and number of sectors of \tess\ data. There is an excess of stars with $\simeq$20-25\,day periods compared to periods outside this range; this is seen in \kepler\ data as well \citep{Reinhold2013}.

\section{\tess\ Rotation Pipeline}\label{sec:prot}

The default \tess\ light curves, produced by the \tess\ Science Processing Operations Center (SPOC), are optimized for exoplanet detection rather than preserving stellar variability \citep{2016SPIE.9913E..3EJ}. To better capture stellar rotation signals, we generated custom light curves using the \texttt{unpopular} package\footnote{\url{https://github.com/soichiro-hattori/unpopular}} \citep{hattoriUnpopularPackageDatadriven2021} package.  \texttt{Unpopular} was specifically designed to retain stellar variability \citep{hattoriUnpopularPackageDatadriven2021} and has been increasingly popular for rotation period studies with \tess\ data \citep[e.g.,][]{Wood2023, Vowell2023}.

For estimating rotation periods, we used a Lomb-Scargle (LS) periodogram. While machine-learning \citep{Santos2021} and autocorrelation \citep{mcquillan_rotation_2014} are popular methods to estimate rotation periods from \kepler\ data, we opted to use LS because it has an easily-accessible and commonly-used implementation within \texttt{astropy} \citep{AstropyCollaboration2022}, it handles unevenly sampled and non-sinusoidal data, and it has higher usage with space-based photometry generally. More generally, LS periodograms have been a standard tool for measuring rotation periods in datasets from \kepler, \ktwo, and \tess\ \citep[e.g.,][]{Reinhold2013, Rebull2016, Mann2016, mannZodiacalExoplanetsTime2017, rampalliThreeK2Campaigns2021}.

\subsection{Building CPM light curves}

Comprehensive details on CPM extraction are available in \citet{hattoriUnpopularPackageDatadriven2021}, with the broader method described in \citet{wangCausalDatadrivenApproach2016} and \citet{wangPixellevelModelEvent2017}. Briefly, \texttt{unpopular} constructs light curves by modeling the pixel response within the target aperture using pixels outside the aperture. This approach assumes that non-aperture and aperture pixels are linked only by non-astrophysical signals (e.g., scattered light and instrumental systematics). The final CPM light curve is produced by subtracting the modeled systematics from the raw aperture light curve, removing instrumental noise without impacting the stellar signal.

For each target (and sector of observations for that target), we downloaded $50 \times 50$ \tess\ FFI cutouts from the Barbara A. Mikulski Archive for Space Telescopes (MAST) using the \texttt{TESScut} interface \citep{2019ascl.soft05007B}. We searched for targets by their \tess\ Input Catalog (TIC) IDs and performed background subtraction on the raw data using the median flux from the 300 dimmest pixels in each cutout. To extract the light curves with \texttt{unpopular}, we applied a one-pixel aperture centered on the target to minimize contamination from background stars, given TESS's large 21\arcsec\ pixels. We selected 100 predictive pixels, excluding a 5x5 pixel region around the target, and used the ``Similar Brightness'' method in \texttt{unpopular}. 

\begin{figure*}[t!]
    \centering
    \includegraphics[width=0.97\textwidth]{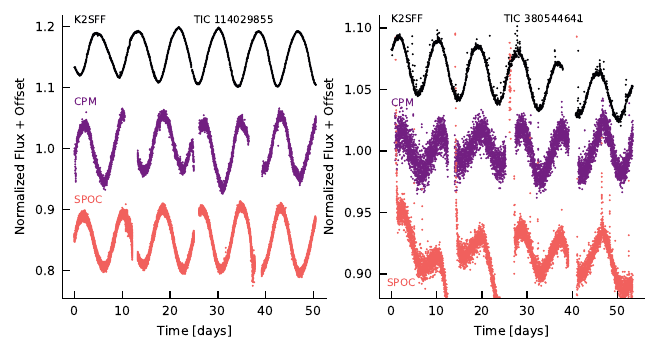}
    \caption{\textbf{Different pipelines have different systematics.} Example stitched-sector CPM light curve (middle) compared to the \ktwo\ curve from K2SFF \citep[top;][]{vanderburg_technique_2014} and the SPOC curve (bottom). Some of the points in the CPM curve missing from SPOC are due to default quality cuts, but CPM also recovers some points where SPOC failed. As can be seen more clearly in the right panel, the rotation signature seen in the \ktwo\ curve is still visible in the \tess\ curve but distorted in the SPOC data.}
    \label{fig:stitched_cpm_vs_k2}
\end{figure*}

For the CPM model, we split each light curve into 100 sections for training and testing, applying an L2 regularization value of 0.1. In this configuration, each section was detrended using a model trained on the remaining 99 sections, with the regularization value setting the strength of the model’s regularization. Example \ktwo, CPM, and SPOC light curves are displayed in Figure~\ref{fig:stitched_cpm_vs_k2}.

This configuration for \texttt{unpopular} has been successfully employed in previous analyses \citep[e.g.,][]{Barber2022, Wood2023}. Our conclusions are also robust against modest variations in this setup (e.g., in the L2 regularization value or aperture size; \citealt{hattoriUnpopularPackageDatadriven2021}). 

\subsection{Quality Cuts}\label{subsec:quality_cuts}

\textit{Contamination ratio} -- To minimize the risk of contamination, we excluded stars with a TIC contamination ratio greater than one. This ratio quantifies the degree of flux blending from nearby sources, with higher values indicating greater contamination \citep{Stassun_2018}. Contamination risk increases with $T$ due to \tess's large pixel size and the higher stellar density at fainter magnitudes.  

Among the 22,986 stars in our sample, 13,506 had contamination ratios below one in the TIC. Most of the difference was from stars with no contamination ratio in the TIC. For example, none of the 545 stars in our sample with $T > 16$ had contamination ratios listed in the TIC. To address this, we recalculated the contamination ratio for each star using the \texttt{tic\_contam.py} script from \cite{2021arXiv210804778P}, increasing the number of stars with contamination ratios below one from 13,506 to 22,403 (97\% of the sample), including 486 stars with $T > 16$.

\medskip

\textit{Removing binaries} -- Unresolved binary companions can introduce periodic signals into the target star's light curve, potentially leading to the recovery of the binary companion's period instead of the target star's, resulting in discrepancies between the RH20 period and our measured period. To mitigate this, we excluded stars with Gaia RUWE $>$ 1.2 and \texttt{non\_single\_star == 1}, both of which are indicative of an unresolved binary \citep[e.g.,][]{Ziegler2020} or extreme youth \citep{Fitton2022}. We note that this cut will not remove most binaries \citep{wood_characterizing_2021}. Among the 22,986 stars in our sample, 5,363 had RUWE $>$ 1.2, and 548 were flagged with \texttt{non\_single\_star == 1}. 

After all cuts, our sample was reduced to 16,752 stars. This is the set of stars used for all subsequent analyses.

\subsection{Estimating \protCPM}\label{sec:protcpm}

To estimate rotation periods from the CPM light curves we used the Lomb-Scargle (LS; \citealp{lombLeastsquaresFrequencyAnalysis1976}, \citealp{scargleStudiesAstronomicalTime1982}) algorithm with the \texttt{nifty-ls} method\footnote{\url{https://github.com/flatironinstitute/nifty-ls}} in \texttt{astropy}. 

We searched for rotation periods across a grid from 0.2 to 20 days, using 100,000 evenly spaced steps. Other parameters for the LS periodogram were kept at their default values. An evenly spaced grid was chosen because the default \texttt{astropy} grid is sparsely sampled at rotation periods $\gtrsim10$ days, which can lead to incorrect period determinations for longer-period rotators. The cutoff at 0.2\,days was applied because stars with rotation periods below 0.2 days are exceptionally rare \citep{Newton2016,Gunther2022} and challenging to detect in the longest cadence (30m) \tess\ data. 

To identify the peak in the periodogram, we located the highest local maximum in period space and adopted this as the rotation period (\protCPM) for further analysis. For stars observed in more than one sector, we independently measured the rotation period in each sector and selected the period with the highest LS power as the final rotation period for that star. Sector-to-sector variations are discussed in Section \ref{subsec:empirical_uncertainties}.

We repeated this analysis for all stars with data from at least two consecutive \tess\ sectors. For stars with multiple consecutive sectors, we analyzed only the first two sectors by stitching them into a single light curve and applying the LS periodogram with the same parameters as described above. We then extended this analysis by stitching all available sectors for each star to create a composite light curve, on which we again applied the LS periodogram. Results are presented in Section \ref{sec:results}.

\section{Results}\label{sec:results}

To begin, we will only consider rotation periods from individual sectors. We will address the results from stitched sectors in Sections \ref{sub:multi_results} and \ref{sec:comp_rel_multi}.

\subsection{Empirical Uncertainties on \protCPM}\label{subsec:empirical_uncertainties}

To estimate true period uncertainties, we used an empirical approach, comparing rotation measurements $P_i$ obtained from multiple datasets (i.e., multiple \tess\ sectors and \ktwo\ campaigns) with a chosen benchmark period $P_{true}$. The relative differences, $(P_i - P_{true}) / P_{true}$, should be normally distributed around zero, with the standard deviation representing the uncertainty on \prot\ (assuming the values are the same across stars within a bin). Assuming $P_{true}$ is the true period and has negligible uncertainty, this method will capture both data-related uncertainties and astrophysical effects, as the light curves are separated by timescales longer than those of spot evolution and other relevant astrophysical changes. 

\begin{figure}
    \centering
    \includegraphics[width=0.48\textwidth]{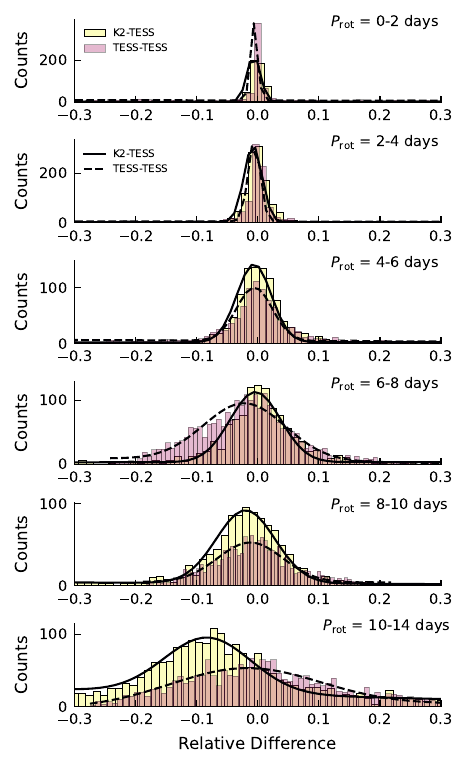}
    \caption{Distribution of the difference between the measured period from \tess\ data and the benchmark period - taken from \ktwo\ data (RH20; yellow) or the best (highest LS power) sector of data from \tess\ (red).  The histograms show the fractional difference between the rotation periods, broken into six period bins. The black lines show fits assuming a Gaussian distribution with a constant offset to account for non-matches (which have random difference).}
    \label{fig:uncertainty_fits}
\end{figure}

We applied this comparison to our \protCPM\ estimates, treating each star’s \protRH\ as $P_{true}$ and each available single-sector \protCPM\ as $P_i$. To analyze the results, we grouped data into 2-day bins in \protRH, increasing to 4 days in the final bin due to lower counts. We show the distribution of each bin in Figure \ref{fig:uncertainty_fits}. 

Within each bin, we performed a least-squares fit assuming a Gaussian plus a constant to the binned distribution of $(P_i - P_{true}) / P_{true}$. We only considered $(P_i - P_{true}) / P_{true}$ values in the $[-0.45,0.45]$ region. The constant and the restricted window are meant to focus on the empirical uncertainties in cases where the period is the correct one, not the effects of aliasing or erroneous measurements (a topic for Section~\ref{sec:rel}). 

\begin{figure*}[t!]
    \centering\includegraphics[width=\textwidth]{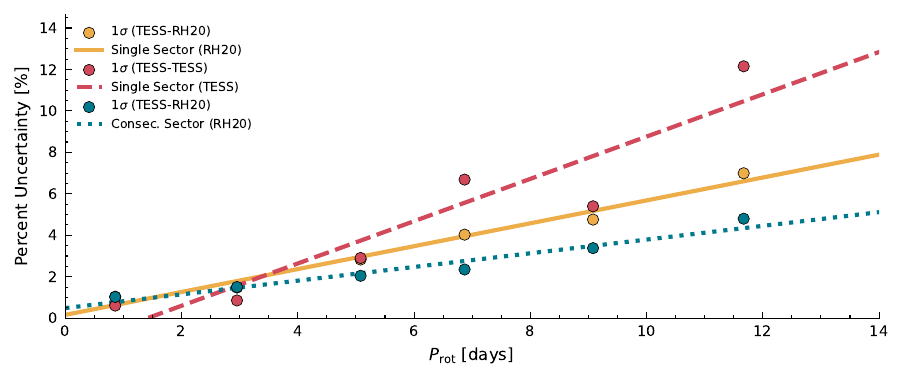}
    \caption{Period uncertainty (\%) on \protCPM\ as a function of \protCPM. The points are derived from the bins shown in Figure~\ref{fig:uncertainty_fits}. The lines show the best-fit linear relation assuming the true period is the RH20 period (yellow) while the red points show the same assuming the true period is \protCPM\ measurement with the highest LS power. The blue line shows the same analysis repeated using periods derived from light curves made from consecutive sectors instead of single-sector light curves and adopting RH20 as the true period.}
    \label{fig:uncertainty_v_prot}
\end{figure*}

As we show in Figure~\ref{fig:uncertainty_v_prot}, the fractional uncertainties scale linearly with \protCPM. Period uncertainties are below 3\% for stars with $P_{\rm rot} < 5$ days, roughly doubling by 12 days. We fit this with a line, which yielded a period uncertainty relation of:
\begin{equation}\label{eq:unc}
    \sigma_{\Prot} (\%) = 0.005577 * \Prot/\rm{days} + 0.001768
\end{equation}
This equation is valid for periods $<$ 12 days. While the analysis includes measurements out to 14\,days, a small fraction of these are 12-14\,days and those that are have low reliability (Section \ref{sec:rel}).

Equation \ref{eq:unc} provides a \textit{fractional} uncertainty. To get the absolute uncertainty on the rotation period, take the output from Equation \ref{eq:unc} and multiply it by the measured rotation period. For example, for a rotation period of 5 days, Equation \ref{eq:unc} predicts a fractional uncertainty of 0.030, or 3\%. Multiplied by 5 days, this gives an uncertainty of 0.15 days.

\begin{figure*}[th!]
    \centering
    \includegraphics[width = 0.97\textwidth]{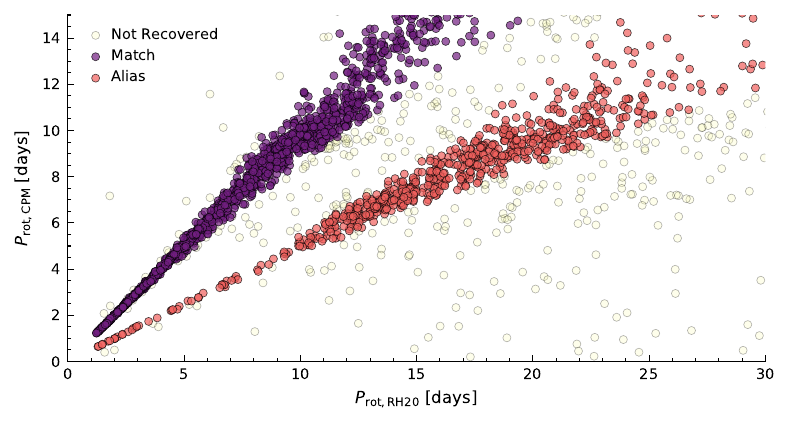}
    \caption{Recovery comparison of \protCPM\ and $P_{\mathrm{rot, \, RH20}}$. We only show points with a Lomb-Scargle power $> 0.1$. Stars considered a match are shown in purple, and those considered a (half-period) alias are shown in red-orange. Yellow points are those that are measured with a Lomb-Scargle power $> 0.1$, but not recovered (not a match or alias).}
    \label{fig:cut_cpm_v_rh20}
\end{figure*}

There is a noticeable bias in the longest \protCPM\ periods, such that  \protCPM\ is, on average, smaller than the equivalent \protRH. The effect can be seen in the bottom 8-10\,day and 10-14\,day bins of Figure~\ref{fig:uncertainty_fits} as well as the unbinned comparison (Figure~\ref{fig:cut_cpm_v_rh20}). This bias is small for periods shorter than 10 days; in the 8-10\,day bin the offset is 1.5\%, which is below the uncertainty (4.8\%), which is the width of the Gaussian fit to the data (the standard deviation). The offset being less than the uncertainty in this bin indicates the bias is not the dominant contributor to the error budget in this bin. This effect can be significant over the sample but not for any individual star.

For the 10-14\,day bin, the effect is $\simeq$10\%, a non-trivial effect even considering the larger uncertainty (6-7\%).

This bias is likely due to the shorter \tess\ observing window ($\sim$30 days) compared to \ktwo\ ($\sim$70 days). For stars with true periods of 10-14\,days, a single \tess\ sector covers only two or three rotations and the LS periodogram will have fewer samples at increasing periods. This would also explain why the effect is not seen when we compare periods from \tess\ data to other periods from \tess\ data (discussed below). 

We repeated our experiment by comparing \protCPM\ against the single-sector \protCPM\ measurement with the highest LS power. The results are shown as purple bins in Figure \ref{fig:uncertainty_fits} and the purple points and line in Figure \ref{fig:uncertainty_v_prot}. 

If the \protRH\ uncertainties are much smaller than those from \tess, the \tess-\tess\ comparison should yield larger uncertainties at all periods. Interestingly, it gave smaller uncertainties for the shortest periods ($<$4 days). This could be due to the 6-hour signal in \ktwo\ light curves \citep[from the roll and thruster fire][]{VanCleve2016, vanderburg_technique_2014}. This would manifest as higher uncertainties in \protRH\ for the fastest rotators. 

The \tess-\tess\ comparison yielded much larger uncertainties in the 6–8 day and 10–14 day ranges. The 10–14 day bin contains the scattered light signal ($\sim$13.7 days), while the 6–8 day bin contains its half-alias. This impacts {\it both} periods used in the comparison, resulting in much higher uncertainties. This also manifests as an asymmetry in the profiles, particularly in the 6-8\,day bin due to a bias towards 6.85 days (half the scattered light signal). 


\subsection{Reliability}\label{sec:rel}

In order to quantify reliability (and completeness in the next section), we needed to quantify if our measured $P_{\mathrm{rot, CPM}}$ matched $P_{\mathrm{rot,RH20}}$. We considered a period a match if $P_{\mathrm{rot,CPM}}$ and $P_{\mathrm{rot,RH20}}$ are within $3\sigma$ of the uncertainty predicted by the linear fit in Equation \ref{eq:unc}, and a similar (proportional) criteria for an alias. Formally: 
\begin{equation}\label{eq:recovery}
    \text{Recovery} = 
    \begin{cases}
        \text{M,} & \left| \frac{ P_{rot,\rm{CPM}} - P_{rot,\rm{RH20}} }{ P_{rot,\rm{RH20}} } \right| < 3\,\sigma_{P_{rot,\rm CPM}} \\[10pt] 
        \text{A,} & \left| \frac{ P_{rot,\rm{CPM}} - \frac{ P_{rot,\rm{RH20}} } { 2 } }{ \frac{ P_{rot,\rm{RH20}} } { 2 } } \right| < 3\,\sigma_{P_{rot,\rm CPM}} \\[10pt]
        \text{NR,} & \text{otherwise}
    \end{cases}
\end{equation}
with M representing a match, A an alias, and NR representing a non-recovery. We use fractional period uncertainties to match our fractional uncertainties from Equation \ref{eq:unc}, and we opted for three standard deviations as this range will include the appropriate spread of recoveries at any \protCPM\ length.

A common source of failures were aliases, i.e., a recovered period that is an integer ratio of the true one. Due to the short observation window of \tess, these are overwhelmingly measurements at half the true period. This effect has been seen in prior studies using \tess\ data as well \citep[e.g.,][]{Kounkel2022}. The number of periods double the true period (and other integer ratios) was consistent with the background of mismatches (see Figure \ref{fig:cut_cpm_v_rh20}), so we defined aliases to only consider the half-period case (Equation~\ref{eq:recovery}). As with the matches, we see a bias for aliases $P_{\mathrm{rot,CPM}}$ to yield shorter periods than $P_{\mathrm{rot,RH20}}$ at $P_{\mathrm{rot,RH20}}\gtrsim8$\,days.


We measured reliability as a function of three parameters: LS power, $T$ magnitude, and the signal-to-noise ratio (SNR). For SNR, we are interested in the ratio of the variability amplitude to that of the random noise and scaled by the number of rotational cycles in the data. We approximate this as:
\begin{equation}\label{eqn:snr}
    \rm SNR = \frac{A_{90-10}}{P2P_{RMS}}\sqrt{N_{cycles}}
\end{equation}
where $\rm A_{90-10}$ is the 90th percentile of the amplitude of the light curve minus the 10th percentile, $\rm P2P_{RMS}$ is the point-to-point variation in the light curve, and $\rm N_{cycles}$ is the number of rotation cycles present in the light curve. Since our light curves have cadences ranging from 20-seconds to 30-minutes, we binned each light curve to a one-hour cadence before calculating the point-to-point variation (so this is a 1-hour SNR). The resulting equation is not a true SNR, which would need to account for other factors like morphology in the light curve. Instead, this is a useful proxy for SNR.

For LS power, we defined reliability as:
\begin{equation}\label{eqn:reliable}
    \text{reliability} = \frac{\text{M($LS_{min}<$ LS power $<LS_{max}$})}{\text{N($LS_{min}<$ LS power $<LS_{max}$)}},
\end{equation}
where M is the number of period matches, and N the total number of stars that satisfy the LS constraint. This represents the fraction of stars with a matching \protCPM\ and LS power in some range divided by all stars in that range. We repeated these calculations using aliases instead of matches, as well as replacing LS power with $T$ magnitude and SNR (Equation~\ref{eqn:snr}) to compute the effect of all parameters on the period reliability.









\begin{figure*}[t]
    \centering
    \includegraphics[width = 0.97\textwidth]{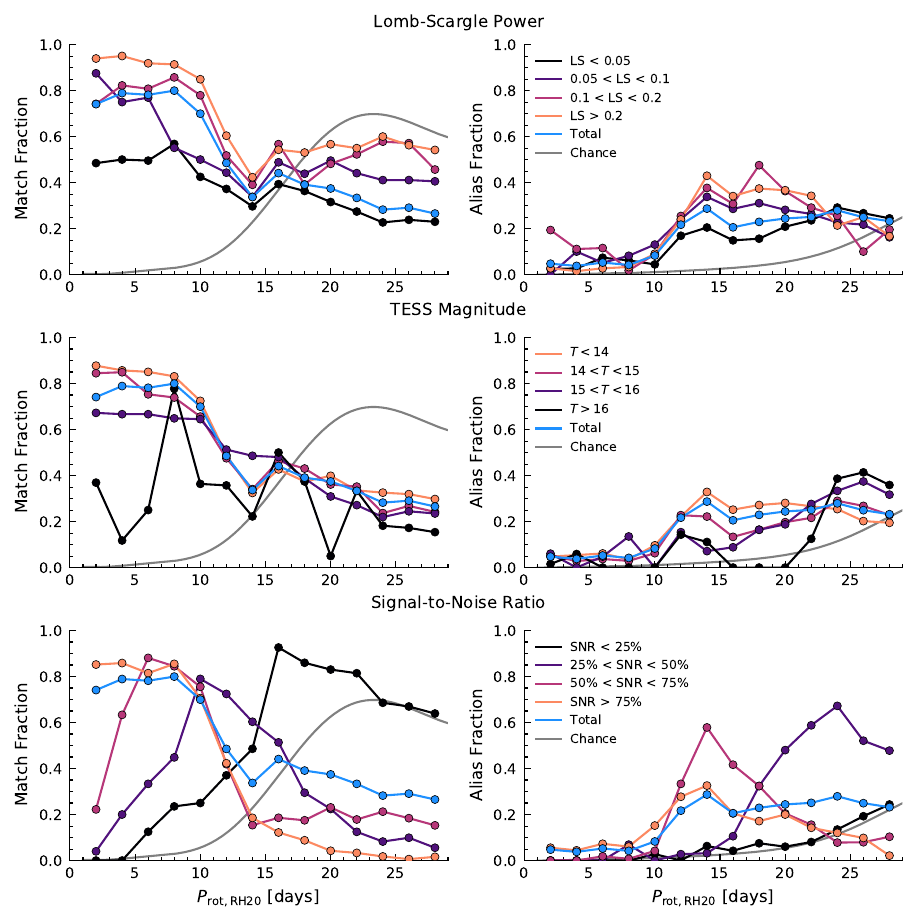}
    \caption{\textbf{Single-sector reliability results.} The left column shows the match fractions for Lomb-Scargle power, $T$, and SNR, while the right column shows the same analysis for aliases. The blue line shows the reliability fraction without cuts, and the gray line represents that chance that a randomly chosen rotation period will match the RH20 period}. Bins with less than three total points have been excluded from these plots.
    \label{fig:single_sector_reliability}
\end{figure*}

Figure~\ref{fig:single_sector_reliability} shows our measurement of reliability as a function of rotation period for varying bins in LS power, \tess\ magnitude, and SNR. Here, we report SNR as percentiles to allow direct comparison between single-sector and consecutive-sector completeness/reliability measurements. Due to the $\sqrt{N_{cycles}}$ term in Equation \ref{eqn:snr}, the consecutive-sector light curves will generally have a higher SNR than the single-sector light curves. As we will see later, while a SNR of 10 might manifest as a strong rotation signal in a single-sector light curve, the same might not be true for a consecutive-sector light curve.


Reliability is high ($>$80\% even with no quality cuts) and relatively flat with increasing period out to $\simeq10$\,days. Reliability drops rapidly around 12\,days (half of a \tess\ sector). There is a corresponding effect seen in the alias fraction: a quick rise between 10 and 15 days. The X-axis is defined as \protRH, meaning $\simeq$30\% of the stars with true periods $\simeq$14\,days are retrieved as 7\,day rotators in \tess\ data (for example). 

The match fraction appears to be $30-60\%$ past 15\,days. However, this is an overestimate driven by definition (Equation~\ref{eq:recovery}). For periods below 8 days, $3\sigma$ corresponds to $\lesssim1$\,day, and hence encompasses a small fraction of the total period space. The odds of a chance match are small ($\lesssim5$\%). By 15 days, $3\sigma$ is about $\pm$4\,days, which would capture 20-40\% of periods in the parent sample by chance. Similarly, at the longest periods, the number of recovered aliases is in-line with the number of aliases that would be recovered by random guessing. We do not consider this to be statistically significant for the same reasons as above --- the uncertainty budget is so large by \prot = 25 days, the number of random matches covers a huge swath of parameter space.

As expected, LS power, $T$, and SNR all impact on the reliability of measured periods. The impact of LS power is particularly strong - curves with the highest LS power have reliability $\gtrsim90$\% even out to 10\,days. In other words, this analysis indicates that if you have a collection of measurements that all have LS power greater than 0.1, more than 80\% of those measurements will be the ``correct'' rotation period until \prot $\approx 8$ days, dropping to 80\% at 10 days, and hitting the sharp drop to less than 50\% by 13 days.


Target brightness has a modest effect for stars $T<15$, after which reliability drops quickly. Periods for the faintest stars ($T>16$) are generally unreliable. SNR appears to have a strong effect; targets with SNR$>10$ are significantly more reliable at periods below 10 days. A lot of this is a consequence of the way SNR is calculated. SNR scales with $\sqrt{N_{cycles}}$ and fast rotators tend to have larger amplitudes \citep{Barber2023}. As a result, fast rotators are overwhelmingly in the high SNR bin and the few that are not tend to be around faint stars. 


One piece that is not captured in Figure~\ref{fig:single_sector_reliability} is why some periods do not match, especially when the LS power is high and the period is short. Of the stars with LS power $>0.2$, \protRH$<$10\,days, and $T<15$, fewer than 5\% are not matches or a half alias. Examination of these stars suggests most of them are periods that landed just outside the requirements set in Equation~\ref{eq:recovery}. Those might be larger-than-typical astrophysical variation (e.g., from differential rotation and spot evolution). Other mismatches in this category are binaries that show two distinct periods in the periodogram and analysis of the \ktwo\ data favored a different period than the \tess\ data. Often such systems have two reported periods in the literature \citep[e.g.,][]{Rebull2018,douglas_k2_2019}. We also re-analyzed the \ktwo\ data for a random subset of these mismatches, and recovered a period consistent with our \tess\ estimate, suggesting the \protRH\ was incorrect for $\lesssim$1\% of the overlap sample. We treat all these cases as 'unreliable' based on our strict definition, but the reliability in this bin is 3-5 percentage points higher if we account for these effects.

\subsection{Completeness}\label{sec:comp}

Separate from the questions of precision and reliability is the question of completeness - a measure of fraction of the sample where you expect to be able to measure \prot. This relates to reliability, but completeness considers stars that you removed because of a given sample cut (in LS power, $T$, or SNR). In general, completeness drops as reliability increases. Ideally, we want cuts on quality that will improve reliability for small costs in completeness. In practice, one selects cuts based on the demands of reliability and completeness set by the science. 

The most robust measure of completeness would be injection/recovery, as is done with exoplanet occurrence calculations \citep[e.g.,][]{Wahhaj2013,Petigura2013,Rizzuto2017}. Unlike for exoplanets, however, we do not have an analytic model of stellar rotation. Instead, we estimate how completeness changes as a function of LS power, $T$, and SNR empirically using our cross-match \ktwo-\tess\ sample. A downside of this approach is that it includes any biases already present in the \ktwo-\tess\ overlap sample, but it is still a good proxy for how completeness changes with \prot\ and data quality cuts. We discuss this issue further in Section~\ref{sec:conclusion}.

We define completeness for LS power as:
\begin{equation}\label{eqn:complete}
    \text{completeness} = \frac{\text{M($\text{LS power} >$ x)}}{\text{N}}.
\end{equation}
As with Equation~\ref{eqn:reliable}, M represents the number of matches satisfying a given criteria. Unlike  Equation~\ref{eqn:reliable}, N is the total number of stars in the \ktwo-\tess\ cross-match sample. Thus, completeness reflects the overall success rate for the full sample (and the number of stars lost in any given cut), while reliability measures the success rate only among stars that meet the detection criteria (i.e., how accurate are the measured rotation periods for stars above the detection threshold).



\begin{figure*}[t]
    \centering
    \includegraphics[width = 0.97\textwidth]{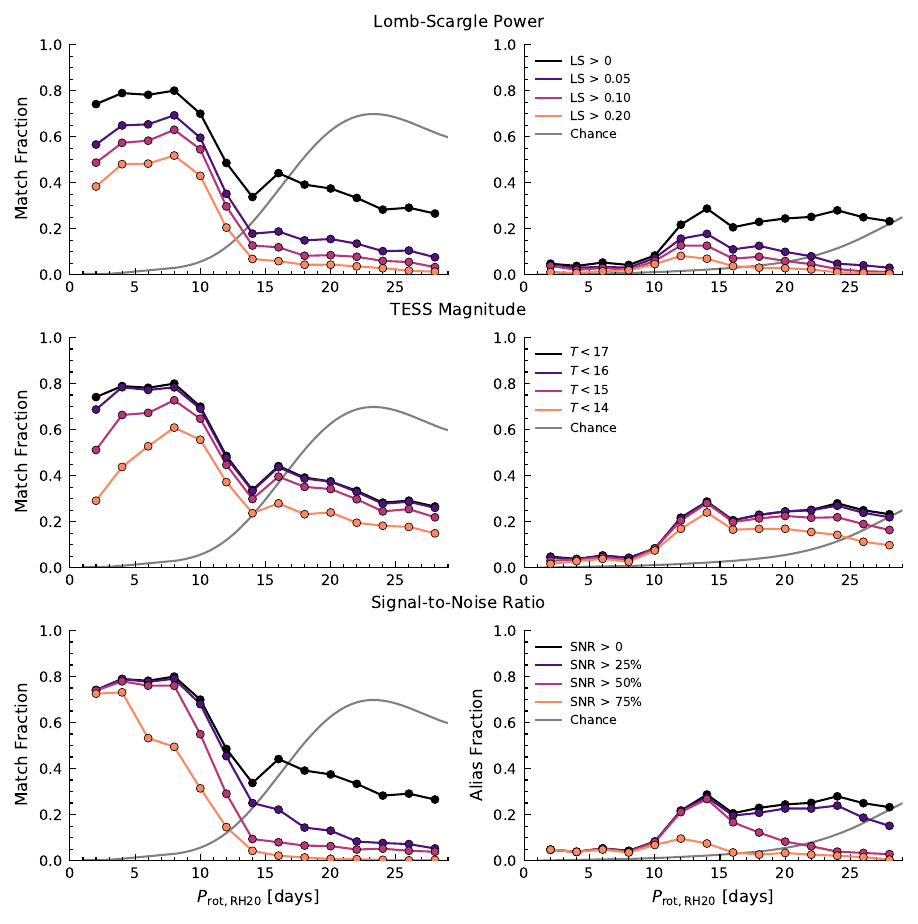}
    \caption{\textbf{Single-sector completeness results.} The left column shows the match fractions for Lomb-Scargle power, $T$, and SNR, while the right column shows the same analysis for aliases. The gray line represents that chance that a randomly chosen rotation period will match the RH20 period. The maximum completeness is $\sim80\%$ because a large number of \tess\ rotation measurements have LS power $<$ 0.2, many of which do not match the true RH20 period. See Section \ref{sec:comp} for a more detailed description of how to interpret completeness results.}
    \label{fig:single_sector_completeness}
\end{figure*} 

Figure \ref{fig:single_sector_completeness} shows how completeness changes with \protRH\ and each of the three tested parameters. As with reliability, completeness drops rapidly at 10-15\,days, a reflection of the low match rate. Below 10\,days, completeness is no higher than $\simeq80\%$. That is, we cannot recover about 20\% of the periods for stars with $<10$\,day rotation periods. The great majority of these 20\% are stars that land in low reliability regimes (e.g., low LS power and faint).

SNR cuts have a small effect on the fastest rotators, as almost all of fast rotators have a high SNR. However, SNR cuts dramatically reduce the number of $\simeq$10\,day rotators. LS power cuts have a more smooth effect, reducing the sample across all periods. The highest LS cut we considered (LS$>0.2$) yielded highly reliable periods below 10\,days (90-100\%), but at a heavy cost. This cut removed more than half the sample. 

Reliability was only weakly impacted by magnitude cuts out to $T\sim$16 and cut out a large fraction of the sample. On the faint end, $T<16$ and $T<17$ are almost identical. This is because there are few stars $T<16$ and few of those are matches (Figure~\ref{fig:single_sector_reliability}). The implication is that magnitude cuts below $T=16$ come at a significant cost in completeness for marginal gains in reliability - nearly all of which are covered by an LS power cut. 

Our completeness analysis is designed to answer the question: Given a sample cut on LS power, $T$, or SNR, how many rotation periods are removed from your sample? For example, if you choose to only consider stars with LS power $>$ 0.2, your measurements will be highly reliable (cf. Figure \ref{fig:single_sector_reliability}) but will have very low completeness (cf. Figure \ref{fig:single_sector_completeness}) because very few stars have a strong enough rotation signal to have LS power $>$ 0.2. In fact, of the 5,723 stars in our sample where our \tess\ period matched the RH20 period, only 1,532 (27\%) had LS power $>$ 0.2.

\subsection{Multi-sector results}\label{sub:multi_results}


The above analysis considers only single \tess\ sectors. The systematics treatment using CPM (as done with \texttt{unpopular}) should preserve astrophysical signals, even over multiple sectors. However, this requires testing; time-dependent signal degradation and long-term systematics may dominate even with careful correction. Here we repeat our analysis after stitching together sectors, exploring the impact on completeness and reliability.

To start, we focus on targets with two consecutive sectors of data. The sectors are combined as described in Section~\ref{sec:protcpm}, and our rotation period estimates are done in the exact same way as for our single-sector analysis. 

As in Section~\ref{subsec:empirical_uncertainties}, we estimate empirical uncertainties by comparing the resulting periods to those from RH20. The resulting comparison is in Figure~\ref{fig:uncertainty_v_prot}. The period uncertainties are lower, particularly at long periods. The gain is about a factor of two by 10\,days, but negligible at periods below 4\,days. At short periods, precision is likely dominated by other systematics (including from the \ktwo\ periods).

\begin{figure*}[t]
    \centering
    \includegraphics[width = 0.97\textwidth]{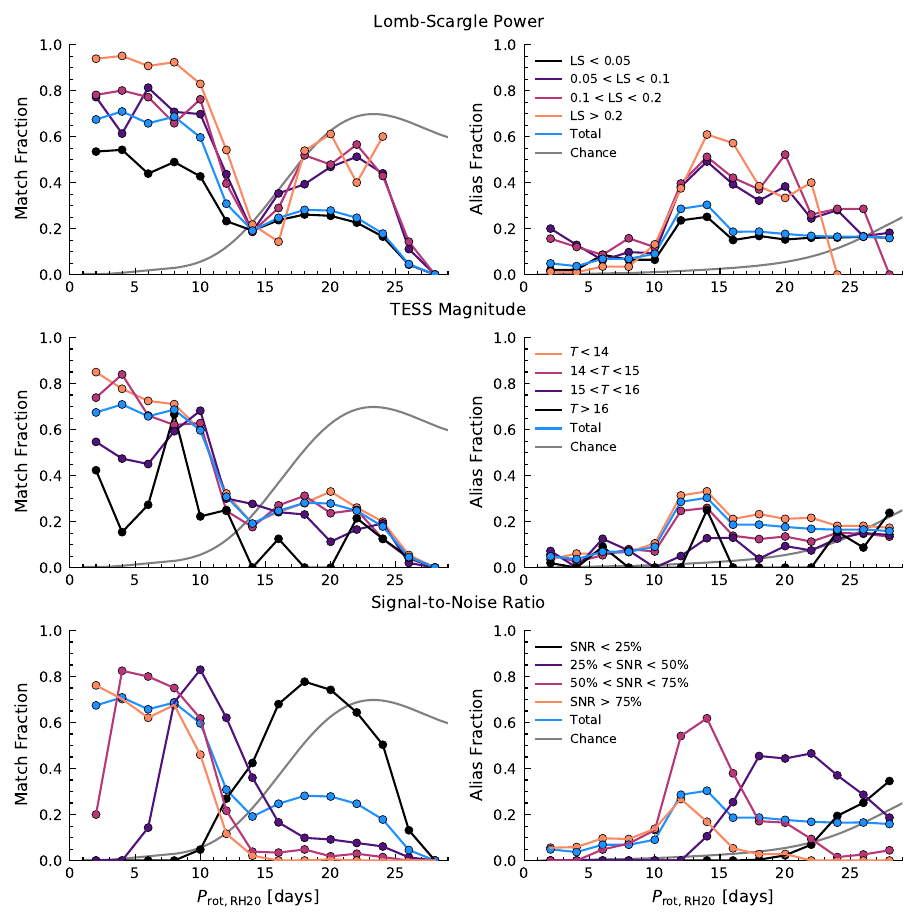}
    \caption{\textbf{Consecutive-sector reliability results.} Same as Figure~\ref{fig:single_sector_reliability} using light curves that were stitched from two consecutive sectors.}
    \label{fig:comp_rel_consec_rel}
\end{figure*}

We show consecutive-sector reliability results in Figure~\ref{fig:comp_rel_consec_rel}. Results below 10\,days are comparable to single-sector results. There are gains in reliability for stars with periods 10-20\,days, but these are mostly targets with the highest LS power (although the reliability is still low). For example, in the LS power $>0.2$ bin, reliability at 15\,days is 20\% for single sectors, and 40\% using consecutive sectors. Other changes are small or consistent with noise.

\begin{figure*}[t]
    \centering
    \includegraphics[width = 0.97\textwidth]{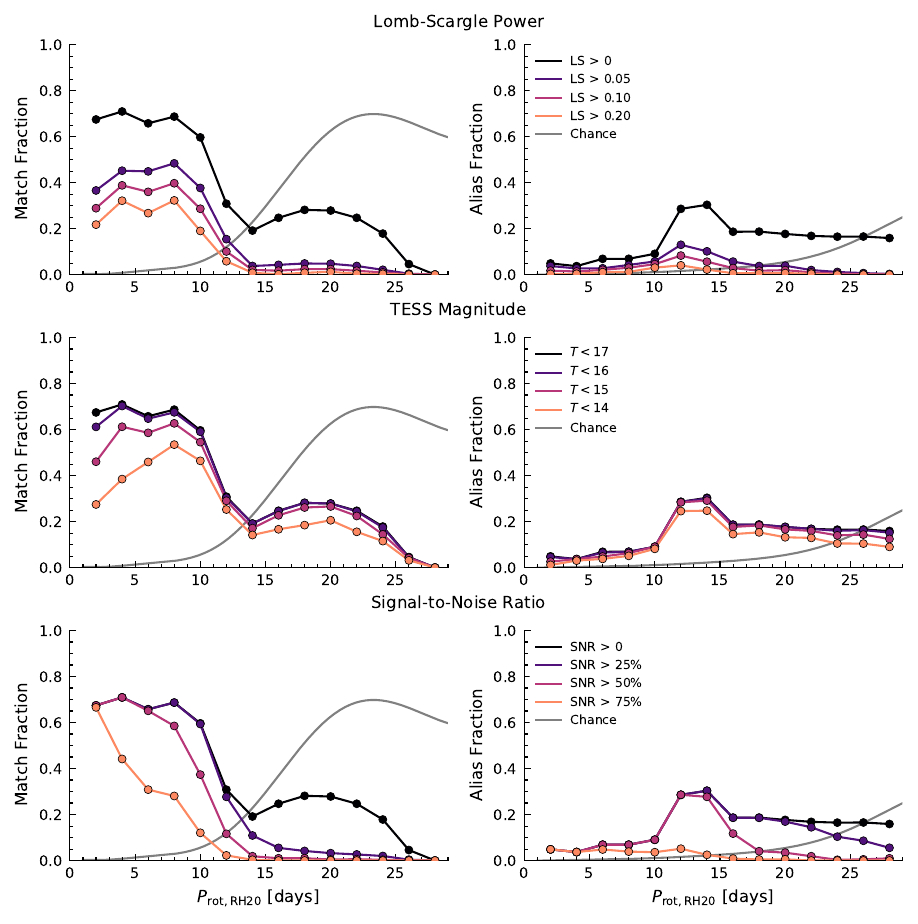}
    \caption{\textbf{Consecutive-sector completeness results.} Same as Figure~\ref{fig:single_sector_completeness} using light curves that were stitched from two consecutive sectors. }
    \label{fig:comp_rel_consec_comp}
\end{figure*} 

We show the completeness results in Figure~\ref{fig:comp_rel_consec_comp}. Similarly to the single-sector analysis in Section \ref{sec:rel}, a LS power cut comes at a significant cost to completeness, with even a modest cut of LS power = 0.05 decreasing completeness by 20\% for \prot\ $<10$ days. The same trend is seen for $T$ and SNR.

\subsection{All-sector analysis}\label{sec:comp_rel_multi}

Our final analysis explores whether the stitching of all available sectors of \tess\ data (consecutive or not) improves the results. Unfortunately, very few stars had more than two consecutive sectors, so this often meant merging light curves with large gaps between the dataset. The net result was only marginal improvement. The period uncertainties were within 1\% of the results from the consecutive sector analysis. Completeness and reliability metrics also showed little change from the consecutive-sector results; most were a {\it decrease} in reliability or a change consistent with random variations.

\section{Application to Stellar Associations}\label{sec:application}

To demonstrate one application of our study, we compiled membership lists for three benchmark associations from the literature: $\alpha$ Per \citep{2023AJ....166...14B}, Pisces-Eridanus \citep{2019AJ....158...77C}, and Group X \citep{2022A&A...657L...3M}. These lists were refined following the criteria in \cite{2023ApJ...947L...3B}, retaining only stars with \texttt{flag\_benchmark\_period == True}. We then used \texttt{unpopular} to generate \tess\ light curves for each star and measured rotation periods using a Lomb-Scargle periodogram, employing the same setup as described in Section \ref{sec:prot}.

We then assign reliability estimates using our \tess-RH20 sample above as a function of the three main parameters: Lomb-Scargle (LS) power, $T$, and SNR. Reliability was calculated using LS power for $\alpha$ Per, $T$ for Pisces-Eridanus, and SNR for Group X (one parameter was used for a given group).  We then select a set of 'nearby' stars within the relevant parameter space. This is always 1-day in period, then $\pm0.05$ around the LS power, $\pm0.5$ mag around the $T$ magnitude, or $\pm2.5$ for the SNR. Any combination of parameters can be used (in addition to period), although for this test we are exploring just one parameter at a time. 

\begin{figure*}[t]
    \centering
    \includegraphics[width = 0.97\textwidth]{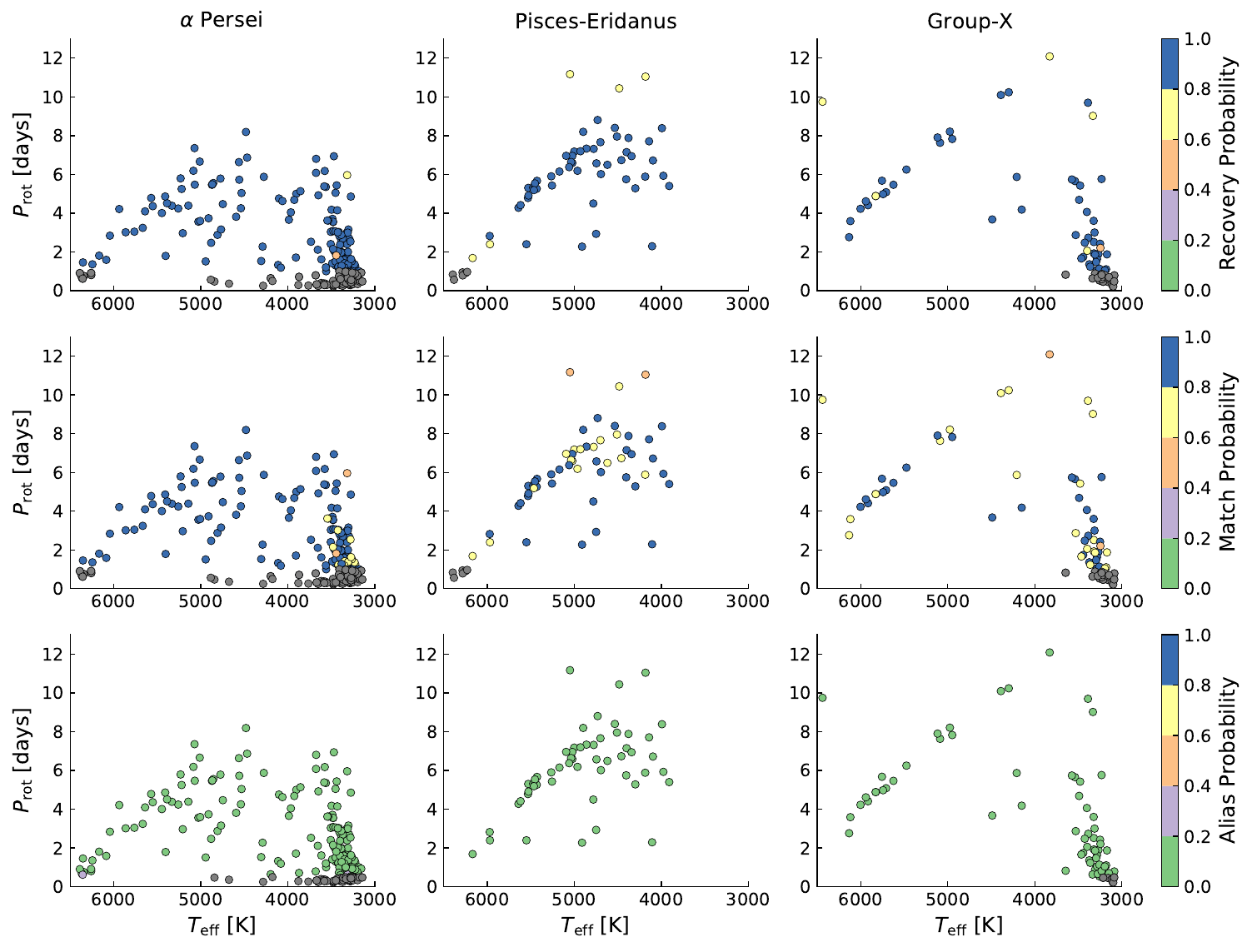}
    \caption{\textbf{Application to young associations.} The recovery, match, and alias probability of rotation periods measured by \tess\ for three benchmark associations: $\alpha$ Persei ($t \sim 80$ Myr), Pisces-Eridanus ($t \sim 130$ Myr), and Group X ($t \sim 300$ Myr). Effective temperatures were estimates from $B_P-R_P$ color using the relation of \citet{}. Match and alias probability were calculated using using the results of our reliability analysis in Section \ref{sec:rel}, with the reliability for $\alpha$ Persei calculated using Lomb-Scargle power, Pisces-Eridanus using $T$, and Group-X using SNR. The recovery probability is the chance that either a match \textit{or} an alias was found. Stars are colored by their probability, with grey points falling in regions where our analysis does not apply.}
    \label{fig:cluster_rotation}
\end{figure*}

Figure \ref{fig:cluster_rotation} presents these results, with each star color-coded by its recovery (top row), match (middle row), and alias (bottom row) probabilities. The recovery probability represents the likelihood that the measured rotation period is either a match or an alias of the true period. Rotation periods shorter than one day are excluded from reliability calculations due to RH20's search limits, and stars with such periods (below one day for matches, 0.5 days for aliases) are shown in gray without probabilities.

Our analysis indicates that the overwhelming majority of stars have high (80-100\%) reliability. Longer-period rotators ($P_{\rm rot} > 10$ days) in Pisces-Eridanus exhibit lower reliability (40\%–60\%) when assessed using $T$; these are also the most discrepant points from the sequence. Overall, we estimate that 11, 7, and 11 rotation measurements are unreliable across $\alpha$ Per, Pisces-Eridanus, and Group X, respectively.


This analysis is somewhat misleading, because our reliability estimates depend on the true period, not the measured one. However, reliability is quite high out to 10-days, which encompasses most of the stars considered here. Although it likely underestimates the number of half aliases (with real periods 10-20\,days) in the sample. One could correct for this by repeating the analysis using the measured period, although this still requires some knowledge of the relative occurrence of stars in a given period bin (i.e., P(\prot)).

We can perform an analogous analysis with completeness. Stars with measured periods likely have higher LS power than their non-measured counterparts (although the effect of the other parameters is less clear). However, assuming measured periods are typical, the number of missing detections (M) is expected to be:
\begin{equation}
    M = \displaystyle\sum\limits_{i=1}^{N}C_i,
\end{equation}
where $C_i$ are the individual completeness estimates for each of $N$ stars with successful measurements. The summed value is effectively the number of 'missed' period measurements for each successful one. In this manner, we estimate there are $17$, $5$, and $14$ members for which we would have failed to measure their rotation period in $\alpha$ Per, Psc-Eri, and Group X, respectively.


\section{Discussion}\label{sec:discussion}

\subsection{Does combining sectors of \tess\ data increase completeness or reliability?}

\begin{figure*}
    \centering
    \includegraphics[width=\textwidth]{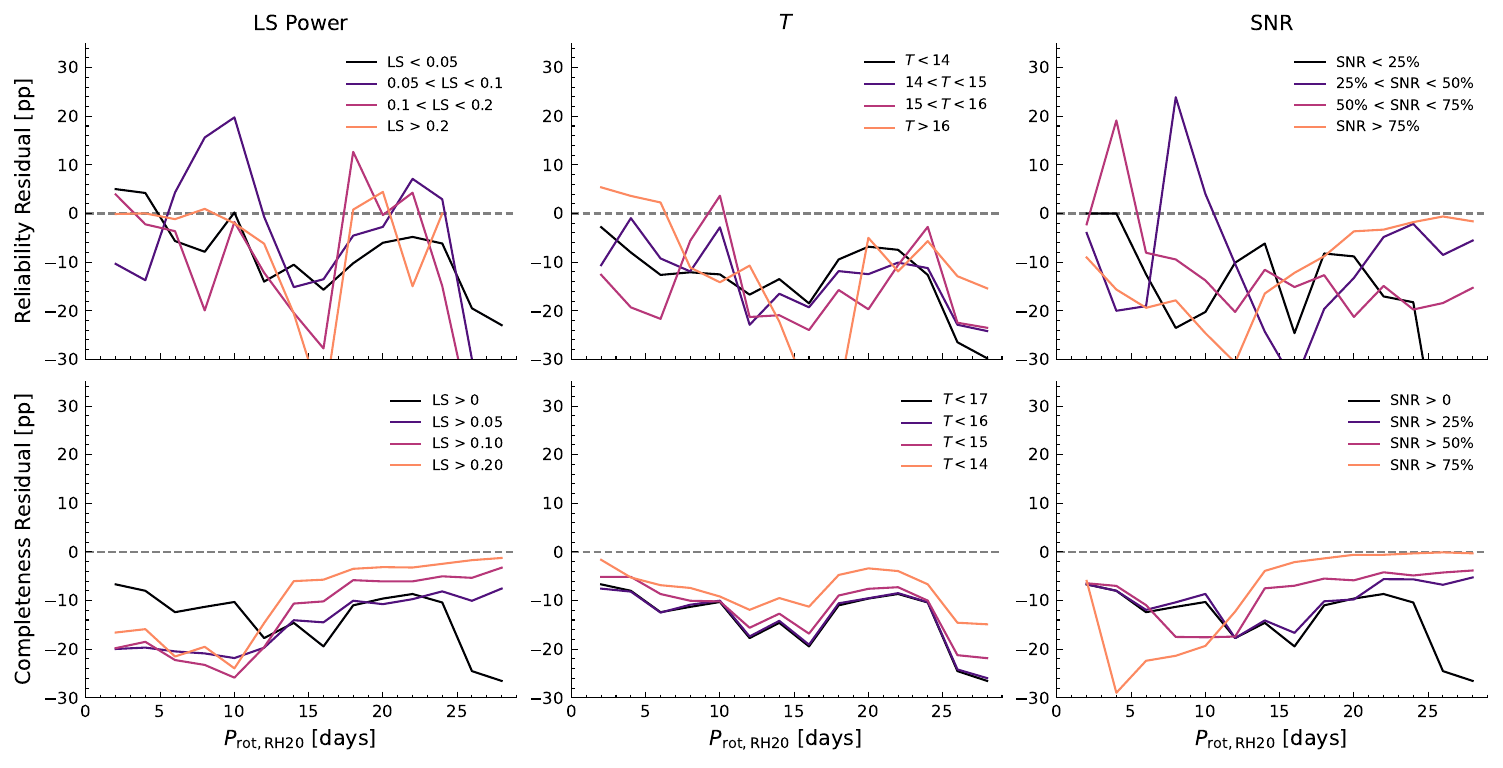}
    \caption{\textbf{Single- and consecutive-sector comparison.} Completeness and reliability of consecutive-sector \tess\ rotation periods compared to single-sector measurements. The two rows are: (1) reliability and (2) completeness of consecutive-sector rotation periods, with residuals showing the consecutive-sector values minus the single-sector values as measured in percentage points (pp).}
    \label{fig:residuals}
\end{figure*}

In principle, longer time baselines can help resolve low-amplitude modulations and better sample slower rotations. Indeed, combined sectors yield lower period uncertainties (Figure~\ref{fig:uncertainty_v_prot}) . One might expect consecutive-sector observations to outperform single-sector data in terms of completeness and reliability. Figure \ref{fig:residuals} illustrates the results by showing the fraction difference between consecutive sector and single-sector results. Positive values mean consecutive-sector measurements outperform single-sector measurements, whereas negative values indicate they underperform.

Generally, consecutive-sector analyses yield 5--10 percentage points {\it lower} reliability and completeness. There are some gains in reliability at lower LS power (0.05--0.10) and SNR (25-50 percentiles) and 5-10\,day periods, suggesting that the longer-baseline does provide advantages in specific cases. Using all available sectors was worse, yielding no significant gains in precision, and additional losses in reliability and completeness. 

Inspection of the results suggest that merging sectors increases both the signal of interest and any instrumental systematics. This also explains why the reliability drop is largest at 12-17\,days and fainter stars, where scattered light is expected to dominate. Merging sectors enhances these systematics because many are common between observations. Changes in spot morphology can happen on timescales of months \citep{rampalli_three_2021}, so the contribution from the star often grows more slowly with increasing number of sectors than other persistent signals.



Although \texttt{CPM} systematic correction is meant to remove instrumental signals while preserving stellar ones, the results suggest this process was not entirely successful. Instead, consecutive-sector measurements should only be used in cases where one is already confident the period is reliable and wants the gain in precision. A better option is to analyze each sector separately and combine the results; simply using the sector of data that yielded the highest LS power provided the highest reliability. An alternative would be to use external ground-based photometry \citep[e.g.,][]{Howard2021} to separate out systematics. Absent that, improvement will require new methods to merge light curves from multiple sectors and techniques to suppress non-astrophysical signals. These results align with other studies that have struggled to measure long-period rotations from consecutive-sectors of \tess\ data \citep{Kounkel2022}. However, more advanced techniques, such as machine-learning techniques \citep{Claytor_2024}, appear to be making progress towards reliably extracting long-period rotation measurements from \tess\ data.


\subsection{Effective Quality cuts}

One goal of our analysis is to identify the quality cuts that maximally improve reliability, minimize spurious detections, and/or maintain high completeness. The three parameters explored here (LS power, $T$ magnitude, and SNR) are similar to or identical to the most frequently applied \citep[e.g.,][]{2021AJ....162..197B, 2023ApJS..268....4F}. 

Our results show the reliability/completeness trade-off. The highest LS power cut ($>$0.2), for example, yields $>90\%$ reliability for periods below 10\,days. However, even among fast rotators, most light curves do not yield such high LS power. So, this cut would catch only $40-50\%$ of the sample. Similar effects are seen from cuts on $T$. The pattern with SNR is more complex, in part because there are few fast-rotating low-SNR targets. The co-dependence of quality cuts is more complicated, as each of the parameters we explored are correlated to each other. 

To facilitate setting cuts specific to a given science case, we provide a code that calculates the completeness and reliability for any given subset of the sample\footnote{\url{https://github.com/awboyle/comp_rel}}. In this way, one can design their cuts around the science requirements or determine period reliability on a per-star basis over the survey. 

\section{Conclusion}\label{sec:conclusion}

We evaluated the precision, completeness, and reliability of stellar rotation measurements derived from \tess\ data. To this end, we used a sample of 23,000 stars with \tess\ light curves and rotation period measurements from \ktwo\ data. By assuming the \ktwo-based periods are highly reliable, we could assess the performance of periods derived from \tess\ light curves as a function of LS power, $T$ magnitude, SNR, and rotation period. 

In terms of precision, we find that:
\begin{itemize}
    \item Periods derived from a simple Lomb-Scargle of \tess\ data are good to 1-2\% for periods below 5 days, increasing roughly linearly to 6\% by 12\,days. 
    \item The precision at periods from 6--8\,days is significantly improved by merging sectors, although the gains are similar to computing each sector separately and combining the resulting measurements. 
    \item \tess-based periods are systematically biased by $\simeq10\%$ (too short) for stars with periods of 10--14\,days. This effect is weak in the 8--10\,day bin ($<3\%$) and not measurable at shorter periods. 
\end{itemize}
Because we compared periods estimated from different instruments and separated by years (over which the stellar signal chances) these estimates cover any instrumental and astrophysical effects.

In terms of reliability, we find that:
\begin{itemize}
    \item Rotation periods derived using a single sector of \tess\ data are $70-80\%$ reliable out to periods of 10 days, even without any quality cuts. Modest cuts can improve this to more than 90\%. 
    \item Beyond 10\,days, reliability drops below 40\%, and by 15\,days assigned periods are only marginally better than random assignment.
    \item LS power is a strong predictor of reliability. The highest LS cut ($>$0.2) gives periods that are reliable 90-95\% of the time out to 10 days. Brightness has a more modest impact on reliability, mostly in the faint end ($T>15$)
    \item Alias detection is minimal ($\sim10$\%) for periods shorter than 5 days but increases to $\sim30$\% for periods of 10 days or more. 
    \item Stitching consecutive or multiple sectors does not enhance reliability; single-sector measurements are generally as effective. 
\end{itemize}

For completeness, our findings indicate:
\begin{itemize}
    \item Maximum completeness is $\sim80\%$, though completeness drops significantly with stricter LS power, $T$, or SNR cuts.
    \item Stitching multiple sectors does not improve completeness.
\end{itemize}


\subsection{Applications to young associations}

One of the goals of developing a probabilistic assessment of each rotation period measurement was to include rotation data into higher-level analyses. Specifically, we are interested in both identifying new members of known associations and using rotation to find entirely new associations. 

A Bayesian selection of association members follows the form:
\begin{equation}
    P(M|\vec{\theta}) = \frac{P(\vec{\theta}|M)P(M)}{P(\vec{\theta})},
\end{equation}
where $P(M|\vec{\theta})$ is the probability that a star is a member given the set of measurement parameters $\vec{\theta}$. In most cases, $\theta$ is a set of kinematic and spatial coordinates (e.g., $UVWXYZ$) that can be compared to a model of the known group(s) and of the Galactic field \citep{rizzuto_multidimensional_2011, malo_banyan_2014}. 

To properly consider rotation as a measurement parameter requires $P(P_{\rm rot}|M)$, the probability of {\it measuring} $P_{\rm rot}$ given that a star is a member. For this, we can use gyrochronology relations \citep[e.g.,][]{bouma_rotation_2021} to build a model of \prot\ for a given star including astrophysical variation. But turning the model into a probability of measuring that value, requires both an estimate of the fraction of unreliable periods and the probability that a period was missed even if it is present (i.e., the reliability and completeness). The remaining missing element is $P(P_{\rm{rot}})$. Although this can be derived from a larger parent population, like \kepler, or from a model of rotational spin-down \citep{matt_spin_2012}.



This framework allows membership algorithms to capture the full uncertainty in the data: if a star’s rotation period is not well-measured (e.g., the reliability is low), the algorithm can degrade its membership probability accordingly. Over many stars, this has the advantage of preventing systematically incorrect rotation periods from skewing cluster properties such as age estimates (through gyrochronology), rotation dispersion, or spin-down rates. Lastly, this method facilitates the selection of more stars further from the group core {\it and} helps remove non-members near a group by chance. 

A separate application is identifying new associations. Many searches for stellar associations have favored the hierarchical clustering algorithm \texttt{HDBSCAN} \citep{mcinnesHdbscanHierarchicalDensity2017}. \texttt{HDBSCAN} enables clustering in an arbitrary number of dimensions, meaning one can add \prot\ as a parameter in addition to the spatial and kinematic parameters used by similar searches \citep{kounkel_untangling_2019, kerr_stars_2021}.

When clustering with rotation as an input parameter, precision and reliability can be used to weight certain points in the clustering metric. As a result, the final cluster assignments will reflect the degree of confidence in each rotation measurement, yielding a cleaner separation of rotational sequences and fewer misclassifications.

\subsection{Limitations and Future work}

A drawback of these conclusions is that we are limited to the set of rotation periods inside the \ktwo\, sample. That is, completeness is only measured with respect to stars with periods below 44\,days, and the comparison set is likely missing many targets at the high end of that period distribution. It also exudes rapidly-rotating stars with weak spot signatures (e.g., some high-temperature stars), although such cases are rare. This has minimal impact on period precision estimates or our measure of reliability, but it means we are overestimating our completeness compared to the true parent population of rotators. 

Equation~\ref{eqn:complete} is therefore more useful as a relative completeness (how it changes with various cuts). Absolute completeness is significantly lower, as there are many stars with periods longer than 40\,days. This could be computed from our measurements using Bayes theorem (P(\prot |LS,$T$,SNR)) given the underlying distribution of rotation periods (P(\prot)). A partial fix for this is to use a benchmark sample with longer-period rotators, like \kepler\ or ground-based surveys with long baselines \citep[e.g.,][]{Newton2016}. 

Our analysis used the Lomb-Scargle periodogram, as it is the most common method for estimating rotation periods from light curves. Over such large samples, machine-learning methods may yield better results \citep{Claytor_2024}. The results here can still be used as the benchmark for testing performance (e.g., test if the new method outperforming a simple Lomb-Scargle). 

A major area for improvement would be methods to stitch the sectors. Right now, combining sectors worsens the output, which is hard to overcome absent a method to absolutely calibrate the fluxes in each sector. One could fit for a flux offset between each sector, and adopt the combination that yields the strongest period. However, this is not computationally practical when running over $>10^4$ stars. Such an approach would also need to be combined with a method to suppress scattered light and other signals not connected to the star. 





\bigskip

The authors would like to thank Luke G. Bouma for his helpful comments. This material is based upon work supported by the National Science Foundation Graduate Research Fellowship Program under Grant No. DGE-2439854. Any opinions, findings, and conclusions or recommendations expressed in this material are those of the authors and do not necessarily reflect the views of the National Science Foundation. 

A.W.B. was partially supported by a grant from NASA's Astrophysics Data Analysis Program (80NSSC24K0619). A.W.M is supported by a grant from NASA's Exoplanet Research program (80NSSC21K0393) and the NSF CAREER program (AST-2143763).

This paper includes data collected by the TESS mission. Funding for the TESS mission is provided by the NASA's Science Mission Directorate. This paper includes data collected by the K2 mission. Funding for the K2 mission is provided by the NASA Science Mission directorate.


%

\vspace{5mm}
\facilities{TESS \citep{2015JATIS...1a4003R, https://doi.org/10.17909/0cp4-2j79}}

\software{Astropy \citep{astropy:2013, astropy:2018, astropy:2022},  
tess-point \citep{2020ascl.soft03001B},
matplotlib \citep{Hunter:2007},
pandas \citep{reback2020pandas, mckinney-proc-scipy-2010},
unpopular \citep{hattoriUnpopularPackageDatadriven2021}
          }



\appendix

\section{What happens if we adopt \kepler\ rotation periods as the benchmark set?}

\begin{figure*}[t]
    \centering
    \includegraphics[width = 0.97\textwidth]{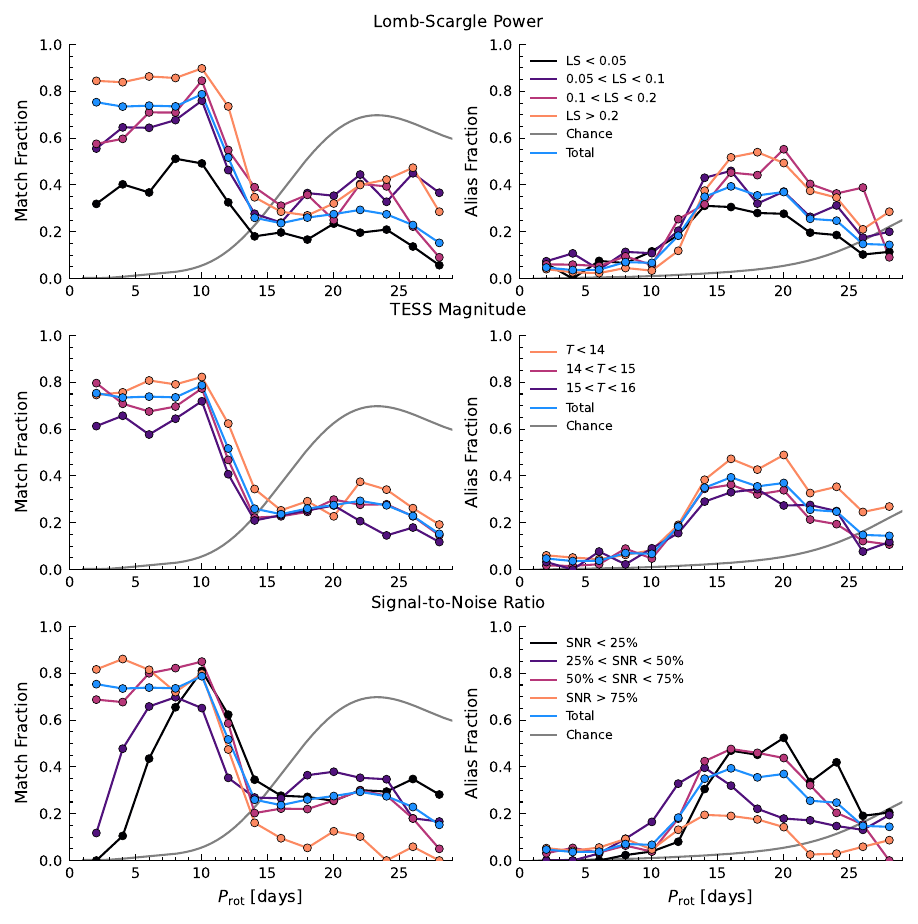}
    \caption{Reliability analysis with \kepler\ rotation periods. Our results show the same trends as in our earlier \ktwo-\tess\ overlap, namely that the reliability of \tess\ rotation periods remains high out to $\sim$10 days and drops sharply thereafter. Aliases increase as rotation period increases.}
    \label{fig:rh13_comparison}
\end{figure*} 

As an additional verification on our results, we repeated our reliability analysis using rotation periods from the original \kepler\ mission as our benchmark set instead of \ktwo\ periods. Specifically, we selected 5,000 stars at random from \citet{2013A&A...560A...4R} (hereafter R13), who measured rotation periods for active \kepler\ stars. This sample is well suited to our comparison: the stars are generally younger and more rapidly rotating, so they should be more amenable to rotation period characterization with \tess\ than a selection of slower rotating field stars.

We applied the same selection criteria to the R13 sample as in our primary analysis: we required RUWE $< 1.2$, $T < 17$, and TIC contamination ratio $<1$. For each star, we generated a \tess\ light curve using \texttt{unpopular} and measured its rotation period with a Lomb–Scargle periodogram, adopting the same settings as before. We then recomputed our reliability metric using the R13 and \tess-based rotation periods.

Figure~\ref{fig:rh13_comparison} shows the results. As with the \ktwo-\tess\ comparison, reliability remains high ($>$80\%) for periods shorter than 10 days, and then falls off rapidly. Beyond 13 days, the recovery rate is worse than random guessing. Due to the R13 sample selection, very few stars have $T > 16$, so we excluded such stars from this analysis to avoid introducing statistical noise. These results verify that the trends we recovered in our earlier analysis with the \ktwo-\tess\ overlap are real.

\section{Receiver Operating Characteristic (ROC) Curves}

\begin{figure*}[t]
    \centering
    \includegraphics[width = 0.97\textwidth]{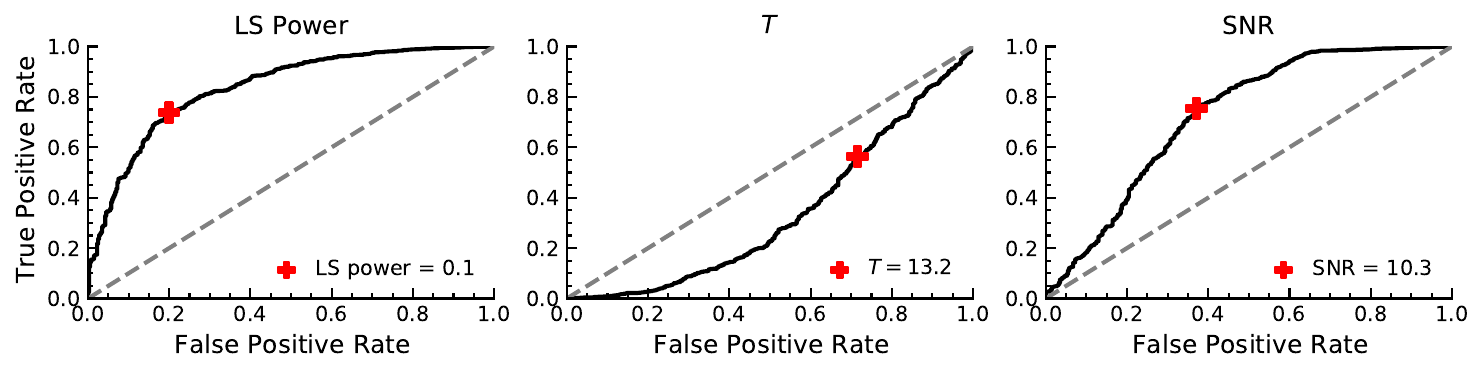}
    \caption{ROC curves for LS power, $T$, and SNR. The gray line represents a model that does no better than guessing at differentiating between true and false positives and is included for reference. The point closest to the top-left corner of each plot --- typically considered the ``ideal'' parameter threshold --- is marked in red and labeled with its value.}
    \label{fig:roc}
\end{figure*} 

As an additional tool for evaluating period reliability, we constructed receiver operating characteristic (ROC) curves for our three parameters (LS power, $T$, and SNR). These curves allow us to assess each parameter's ability to distinguish between rotation period measurements that agree with the benchmark \ktwo\ periods and those that do not. To compute ROC curves, we first classify each measurement as a true positive, false positive, true negative, or false negative. These categories are defined as follows:


\begin{itemize}
    \item \textit{True Positive (TP)} --- A rotation period from a star that passes a given threshold (e.g., LS power $>$0.2) and for which the \ktwo\ and \tess\ periods \textit{agree} within 3$\sigma$.
    \item \textit{False Positive (FP)} --- A star that passes the threshold but for which the \tess\ and \ktwo\ periods \textit{do not agree} within 3$\sigma$.
    \item \textit{True Negative (TN)} ---  A star that falls below the threshold and for which the periods \textit{do not agree}.
    \item \textit{False Negative (FN)} --- A star that falls below the threshold and for which the \ktwo\ and \tess\ periods \textit{agree}.
\end{itemize}

The true positive rate (TPR) and false positive rate (FPR) are defined as:

\begin{equation}
    TPR = \frac{TP}{TP + FN}; \quad
    FPR = \frac{FP}{FP + TN}
\end{equation}


A ROC curve is then made by repeatedly varying a parameter threshold and calculating the TPR and FPR at each instance. A perfect model would follow a path that rises vertically to TPR = 1 before turning horizontally to the upper-right corner (FPR = 1). A diagonal line represents random guessing. The optimal parameter value typically corresponds to the point on the curve closest to the top-left corner.

Figure~\ref{fig:roc} shows the resulting ROC curves for LS power, $T$, and SNR. For each, we mark the point closest to the ideal upper-left corner and annotate the corresponding threshold. We find:

\begin{itemize}
    \item LS power performs best at differentiating true positives and false positives, with the ROC curve significantly above the diagonal and an optimal threshold near LS $\approx 0.1$.
    \item SNR also performs well, with an optimal threshold near SNR $\approx 10$.
    \item $T$ performs poorly: its curve lies below the diagonal, indicating that it is worse than random at separating true positive and false positive periods in our sample.
\end{itemize}

These results are consistent with our earlier reliability analysis and confirm that LS power and SNR are the most informative metrics for screening reliable rotation periods. $T$ is a poor discriminator when used in isolation --- likely because it is only indirectly tied to signal quality, whereas LS power and SNR are more direct measures of the strength of a rotation signal.

We caution that these ROC curves are specific to the \ktwo-\tess\ overlap sample. The optimal thresholds may differ for other stellar populations (e.g., young clusters with faster rotators or higher amplitudes). Nonetheless, this analysis offers a quantitative framework for choosing quality cuts that balance reliability and completeness.



\bibliography{gaia_plx, mann, PAPER-Prot_Bayes, references}{}
\bibliographystyle{aasjournal}



\end{document}